\documentclass{iau} 
\usepackage{graphicx}
\usepackage{amsmath}
\usepackage{amssymb} 

\title[IAU 363. Observability of isolated neutron stars at SRG/eROSITA] 
{Observability of isolated neutron stars \\ at SRG/eROSITA}

\author[A. D. Khokhriakova, A. V. Biryukov and S. B. Popov]
{Alena D. Khokhriakova$^{1,2,3}$,
 Anton V. Biryukov$^{2,4}$
 \and Sergei B. Popov$^{1,2}$}

\affiliation{$^1$Faculty of Physics, Moscow State University, Moscow, Russia
\\
$^2$Sternberg Astronomical Institute, Moscow State University, Moscow, Russia
\\
$^3$``Basis'' foundation fellow, Moscow, Russia
\\
$^4$Kazan Federal University, Kazan, Russia
\\email: {\tt alenahohryakova@gmail.com}}

\pubyear{2022}
\volume{363}  
\jname{Neutron Star Astrophysics at the Crossroads:
Magnetars and the Multimessenger Revolution}
\editors{E. Troja \& M. Baring, eds.}
\begin{document}

\maketitle

\begin{abstract}
A four-year sky survey with the use of the eROSITA telescope on board the Spektr-RG observatory 
will provide the best coverage in the soft (0.5–2 keV) and standard (2–10 keV) X-ray ranges, both in terms of sensitivity and angular resolution. We have analysed the possibility of detecting various types of isolated neutron stars with eROSITA. Among already known objects, eROSITA will be able to detect more than 160 pulsars, 21 magnetars, 7 central compact objects, all seven sources of the Magnificent Seven, and two other X-ray isolated neutron stars during the four-year survey mission. 
\keywords{Stars: neutron, pulsars: general, X-rays: stars}
\end{abstract}

\firstsection 
\section{Introduction}

In the first approximation, the modern X-ray astronomy requires two types of instruments.  Studies of individual sources are better performed using instruments with a small field of view that allow achieving low fluxes and high angular resolution (e.g., XMM-Newton and Chandra). 
For  studies of sufficiently large and possibly uniform samples survey observations are usually required (e.g., ROSAT). 

 The statistics indicate that the number of NSs in our Galaxy should be on the order of $\sim 10^9$ (e.g., \cite[Treves et al. 2000]{8}). At the present time, more than 3000
 isolated NSs are known. Below we consider the possibility of detecting both thermal and magnetospheric X-ray emission from isolated neutron stars (NSs) of various types using the eROSITA telescope on board the Spektr-RG (SRG) satellite. The detailed analysis can be found in \cite{2021ARep...65..615K}.

The majority of isolated NSs currently observed in X-rays are radio pulsars with non-thermal magnetospheric X-ray emission 
(see the ATNF catalog\footnote{ Online catalog of ATNF pulsars, see https://www.atnf.csiro.au/research/pulsar/psrcat/}, \cite{6}). Thermal X-ray radiation is detected only from several dozen objects (\cite[Potekhin
et al. 2020]{7}). The most noticeable in the X-ray range are young isolated NSs, which still have a high surface temperature ($\sim 10^6$ K) and/or a powerful magnetospheric radiation associated, ultimately, with the loss of rotational energy. Old isolated objects are too cold for their thermal radiation to be detected; in addition, they have either already slowed down their rotation significantly or/and their magnetic field has decreased, which makes the magnetospheric radiation mechanism ineffective. However, theoretically they can be detected if they accrete enough matter from the interstellar medium (\cite[Ostriker et al. 1970]{10}, \cite[Shvartsman 1971]{11}).



A four-year survey with eROSITA will provide the best coverage in the soft (0.5–2 keV) and standard (2–10 keV) X-ray ranges both in terms of sensitivity and angular resolution (see detailed description in \cite{12}). In addition, the M. Pavlinsky ART-XC telescope installed on board the SRG will also perform an all-sky survey in the standard X-ray range. This will be very important for studying various subpopulations of NSs.


 
\section{Detectability of known isolated NSs by eROSITA}

We parametrize detectability of NSs by the limiting count rate $F_\mathrm{x,min}$. Following analysis in \cite{2021ARep...65..615K} we assume it to be equal to 0.01 count per second. 
In the following subsections we present our results on different types of NSs detectable by eROSITA.

\subsection{Pulsars}

Pulsars (classical and millisecond) and rotating radio transients (RRATs) make up the bulk of the population of known isolated NSs. In total, the ATNF
catalog (\cite[Manchester et al. 2005]{6}) at the time of this writing contains more than $3000$ radio pulsars and RRATs.
To analyse the possibility of detecting X-rays from pulsars we construct a spectral model of such radiation. The spectrum is assumed to be the sum of thermal and power-law spectra.


For the thermal part, we calculate surface redshifted temperature using a method described in \cite{2021ARep...65..615K}. The surface temperature is estimated through interior temperature, and for the latter we use an analytical approximation (see \cite[Igoshev \& Popov 2018]{21}) of cooling curves from \cite{22}.

For the power-law part, we adopt the photon index $\Gamma = 1.7$ (\cite[Becker 2009]{9}). To normalize this spectral component, 
we use the X-ray luminosity, which is a sum of two terms. In the hard range (2–10 keV), we calculate the luminosity following  \cite{25}, and in the soft range (0.1–2 keV) --- in accordance with \cite{9}.

To calculate the photon detection rate at eROSITA, we use the technique described in detail in \cite{26}. 
The number of detected photons per second is:
\begin{equation}
\dot{N}_\mathrm{detected} = \int_{E_1}^{E_2} {\frac{(4 \pi R^2 \pi B_\mathrm{E} (T,E) E^{-1}+ C E^{- \Gamma})  e^{- \sigma N_\mathrm{H}}  S_\mathrm{eff}(E) dE} {4 \pi d^2}},
\label{eq:N-phot}
\end{equation}
where $E$ is the photon energy in keV,
 $E_1$, $E_2$ are the limits of the telescope sensitivity,
$S_\mathrm{eff}(E)$ is the dependence of the effective area of eROSITA on the photon energy,
$C$ is the normalization constant,
$B_\mathrm{E} (T,E)$ is the Planck function,
$d$ is the distance to the source,
$N_\mathrm{H}$ is the column density of hydrogen atoms.

The number of pulsars with the expected photon count rate above the threshold value $F_\mathrm{x} = 0.01$ s$^{-1}$ turned out to be 162. This number includes both classical and millisecond pulsars.
At the same time it is unlikely that any of known RRATs will be reliably detected due to their magnetospheric emission.



\subsection{Other types of NSs}

{\underline{\it Magnetars}} are young neutron stars with a strong magnetic field ($\sim 10^{14} - 10^{15}$ G); their observational manifestations are associated with the magnetic field energy dissipation.
The luminosity and spectrum of a magnetar depend on the phase of activity. The McGill Online Magnetar Catalog\footnote{ http://www.physics.mcgill.ca/~pulsar/magnetar/main.html} contains spectral characteristics related to the quiet state for 30 magnetars and candidates (\cite[Olausen \& Kaspi 2014]{46}). 
According to our detection criterion $F_\mathrm{x} > 0.01$ s$^{–1}$, eROSITA will be able to detect 21 known magnetars in this state.

{\underline{\it Central compact objects}} (CCOs) are sources of soft thermal X-ray radiation located in young (0.3–7 thousand years old) supernova remnants (\cite[De Luca 2017]{47}).
We apply the blackbody (BB) absorbed spectrum model to most of the objects. For two CCOs
(1E 0102.2–7219 and 1WGA J1713.4–3949), the
BB+BB model is used, and for XMMU J172054.5–
372652, the sum of BB and power-law spectra
(BB+PL). According to our detection criterion
$F_\mathrm{x} > 0.01$ s$^{–1}$, eROSITA will be able to detect seven of the known CCOs.

{\underline{\it The Magnificent Seven}} are close isolated NSs with thermal radiation. Their age is approximately $10^5$ --- $10^6$ years. The surface temperatures responsible for the observed X-ray emission range from $\sim$ 40 to $\sim$100 eV.
Similar to the other types of NSs, we analyse the observability of these objects using spectral parameters from the literature (see \cite[Khokhryakova et al. 2021]{2021ARep...65..615K}). 
For all objects of this type, we use thermal absorbed spectra. 
All the stars of the Magnificent Seven can be observed by eROSITA.

\section{Search criteria for new sources}

In addition to observations of already known sources, eROSITA will detect new isolated NSs. 
In a recent study by \cite{62}, the population synthesis of cooling NSs was used to predict the observability of such sources in the eROSITA four-year survey. Depending on the filter configuration, simulations yield 85 to 95 isolated thermal NSs. 
There are two main approaches to the identification of new objects. The first approach
implies cross-correlation, i.e. the use of data from different instruments in different parts of the spectrum.
In particular, NSs are distinguished by a high
ratio between soft X-ray and optical fluxes, $f_\text{X} / f_\text{opt}$ (\cite[Schwope et al. 1999]{72}). 
The second, generally less effective approach is based solely on X-ray data.

From X-ray observations we can define a simple quantity which very roughly characterizes spectral properties of sources.
Let us introduce the hardness for eROSITA:
\begin{equation}
    \text{HR}_\mathrm{eR} = \frac{(1 - 10 \text{ keV}) - (0.1 - 1 \text{ keV})}{(1 - 10 \text{ keV}) + (0.1 - 1 \text{ keV})}.
\end{equation}
The parentheses indicate the fluxes in the corresponding energy ranges. This value can be quite easily obtained from observations. At the same time, it allows one to separate different classes of objects. In particular, on the base of X-ray observations thermally emitting isolated NSs can be confused with active galactic nuclei (AGN).
We estimate the hardness for AGNs at various redshifts $z$ under the assumption that the spectrum is a power-law with the photon spectrum indices $\Gamma = 1.7$ and 1.9. It can be seen from the Fig.~1 that, in accordance with the conclusions of \cite{72}, NSs with thermal radiation are well separated from AGNs (neglecting noticeable absorption in the immediate vicinity of AGNs). At the same time, ordinary radio pulsars and magnetars with power-law spectra turn out to be ``mixed'' with AGNs; in this case, additional criteria are needed to distinguish between different types of the sources.

\begin{figure}[b]
\begin{center}
 \includegraphics[width=3.7in]{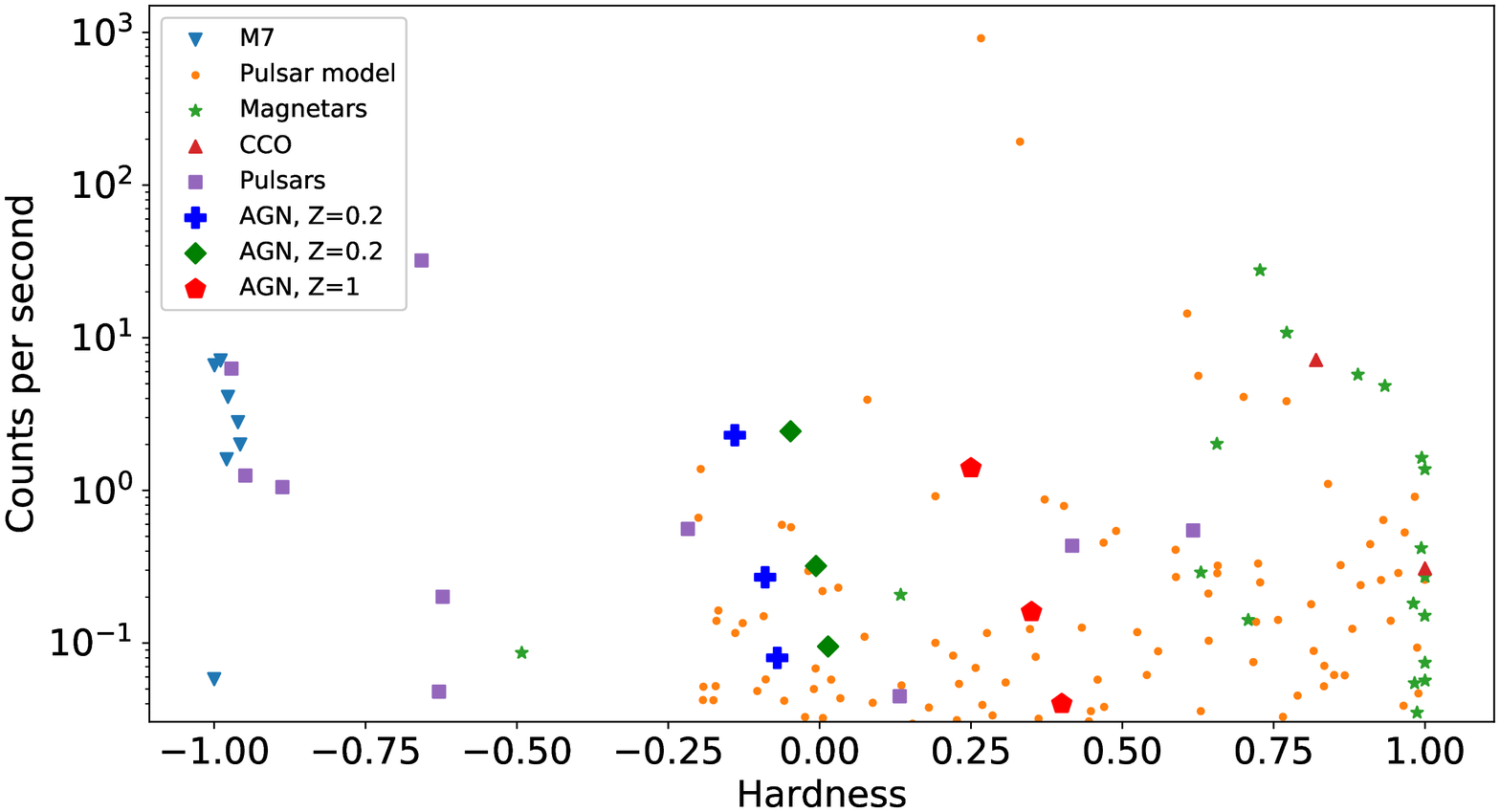} 
 \caption{Hardness versus counts per second diagram for various types of neutron stars that eROSITA can detect. The orange dots are the ATNF pulsars whose X-ray emission was modeled using the model described in Section 2.1. The purple squares represent
pulsars with known thermal X-rays. The green diamonds represent AGNs with a photon index $\Gamma = 1.7$, and blue pluses and red pentagons are AGNs with $\Gamma = 1.9$.}
  \label{fig1}
\end{center}
\end{figure}

\section{Conclusions}

The study of NSs is a rapidly developing field of astrophysics. However, there are rather few currently known isolated NSs emitting in the X-ray range. For this reason, their study and the discovery of new objects of this type is very important.
We have studied the observability of various types of isolated NSs at SRG/eROSITA. Modeling and calculation of the spectra of radio pulsars, RRATs, magnetars, CCOs, and the Magnificent Seven have been carried out. The main results are as follows: among the already known objects, eROSITA will be able to detect $\sim$160 pulsars, 21 magnetars, 7 compact central objects, the Magnificent Seven, and 2 other isolated X-ray NSs over a four-year mission. 
In addition, we expect that eROSITA will be able to detect accreting NSs, as well as discover new cooling NSs and magnetars.

\acknowledgements
SP is supported by the Ministry of Science and Higher Education of Russian Federation under the contract 075-15-2020-778 in the framework of the Large Scientific Projects program within the national project ``Science''.


\begin{thebibliography}{}

\bibitem[Becker (2009)]{9}
{Becker, W.} 2009, {\it Ap\&SSL}, 357, 91

\bibitem[De Luca (2017)]{47} {De Luca, A.} 2017, {\it Journal of Physics Conference Series}, 932, 012006

\bibitem[Igoshev \& Popov (2018)]{21}
{Igoshev, A.P. \& Popov, S.B.} 2018, {\it MNRAS}, 473, 3204

\bibitem[Khokhryakova, Biryukov, \& Popov (2021)]{2021ARep...65..615K} 
{Khokhryakova, A.~D., Biryukov, A.~V., \& Popov, S.~B.} 2021, \textit{Astronomy Reports}, 65, 615

\bibitem[Khokhryakova, Lyapina, \& Popov (2019)]{26}
{Khokhryakova, A.D., Lyapina, D.A., \& Popov, S.B.} 2019, {\it Astronomy Letters}, 45, 120

\bibitem[Manchester \etal\ (2005)]{6}
{Manchester, R.N., Hobbs, G.B., Teoh, A., \& Hobbs, M.} 2005, {\it AJ}, {129}, 1993

\bibitem[Merloni \etal\ (2012)]{12}
{Merloni, A., Predehl, P., Becker, W., B{\"o}hringer, H., Boller, T., Brunner, H. \etal} 2012, {\it arXiv e-prints}, arXiv:1209.3114.

\bibitem[Olausen \& Kaspi (2014)]{46} {Olausen, S.A. \& Kaspi, V.M.} 2014, {\it Astrophys. J. Suppl.}, { 212}, 6

\bibitem[Ostriker, Rees, \& Silk (1970)]{10}
{Ostriker, J.P., Rees, M.J., \& Silk, J.} 1970, {\it Astrophysical Letters}, {6}, 179

\bibitem[Pires, Schwope, \& Motch (2017)]{62}
{Pires, A.M., Schwope, A.D., \& Motch, C.} 2017, {\it Astron. Nachr.}, {338}, 213

\bibitem[Possenti \etal\ (2002)]{25}
{Possenti, A., Cerutti, R., Colpi, M., \& Mereghetti, S.} 2002, {\it A\&A}, { 387}, 993

\bibitem[Potekhin, Chabrier, \& Yakovlev (1997)]{23}
{Potekhin, A.Y., Chabrier, G., \& Yakovlev, D.G.} 1997, {\it A\&A}, {323}, 415

\bibitem[Potekhin \etal\ (2020)]{7}
{Potekhin, A.Y., Zyuzin, D.A., Yakovlev, D.G., Beznogov, M.V., \& Shibanov, Y.A.} 2020, {\it MNRAS}, { 496}, 5052

\bibitem[Schwope \etal\ (1999)]{72}
{Schwope, A.D., Hasinger, G., Schwarz, R., Haberl, F., \& Schmidt, M.} 1999, {\it A\&A}, 341, L51.

\bibitem[Shternin \etal\ (2011)]{22}
{Shternin, P.S., Yakovlev, D.G., Heinke, C.O., Ho, W.C.G., \& Patnaude, D.J.} 2011, {\it MNRAS},  412, L108

\bibitem[Shvartsman (1971)]{11}
{Shvartsman, V.G.} 1971, {\it SvA},  14, 662

\bibitem[Treves \etal\ (2000)]{8}
{Treves, A., Turolla, R., Zane, S., \& Colpi, M.} 2000, {\it PASP}, 112, 297

\end{thebibliography}
\end{document}